\documentstyle[aps,preprint,epsf]{revtex}
\tightenlines
\newcommand{\beq}{\begin{equation}}
\newcommand{\eeq}{\end	{equation}}

\begin{document}


\title{
EMC effect from the QMC and the QHD descriptions
}
\author{
R. Aguirre \footnote{
Electronic address: aguirre@venus.fisica.unlp.edu.ar .}
and M. Schvellinger
\footnote{
Electronic address: martin@venus.fisica.unlp.edu.ar . }}

\address{
Department of Physics, Universidad Nacional de La Plata,
C.C. 67 (1900) La Plata, Argentina.
}

\maketitle

\begin{abstract}

The density dependence of the bag parameters is studied in a
framework which links the Quark-Meson Coupling model and the
field theory of hadrons for nuclear matter description.
The EMC effect is treated in the dynamical rescaling hypothesis
by using the density dependent bag radius and the Local
Density Approximation. Our results are in good agreement with
the expected values of the rescaling parameter.

PACS number(s): 21.65.+f;12.39.Ba;24.85.+p

\end{abstract}

\section{INTRODUCTION}

Ultrarelativistic heavy-ion experiments have reported information
on the influence of the subnuclear degrees of freedom on the
nuclear matter properties \cite{DRUK,BROWN}. The search for
signals of the formation of hot nuclear matter and the
quark-gluon plasma \cite{CPS} has motivated the study of the
density dependence of the hadron masses in the nuclear medium.

Several theoretical approaches have been developed to investigate
the hadron
properties, among them the Quantum Hadrodynamics (QHD) and the
Quark-Meson Coupling model (QMC) use meson fields to propagate the
nuclear interaction.
In the first case structureless baryons are coupled
to mesons. This subject has reached interest since the pionnering
work of Walecka \cite{WALECKA83}. The phenomenology of nuclear
matter and finite nuclei has been studied in QHD with successful
results \cite{SEROT}-\cite{FINITE}. The original Walecka model has
been extended with an additional polynomic potential \cite{BOGUTA}
and non-polynomic interactions \cite{ZM}--\cite{FELD}. In this work
we shall include under the name of Quantum Hadrodynamics all
(the renormalizable and the non-renormalizable) hadronic models.

On the other hand, in the QMC model an explicit hadron substructure
is taken into account. This model is based on a mean field
description where the nucleons are treated as non-overlapping MIT
bags confining the quarks inside them. The model was first
developed by Guichon \cite{GUICHON}. Refinements and applications
have been done \cite{SAITO51}--\cite{CMC}, including the correction
of the center of mass motion \cite{HEPS}. The variation of hadron
masses and matter properties in the nuclear medium \cite{SAITO51}
as well as in finite nuclei \cite{HEPS} have been reported.

A clear relationship between the QMC model and the Walecka model
has been stablished by Saito and Thomas \cite{SAITO200}. In this
work we assume the validity of the nuclear matter predictions
derived from different effective hadronic lagrangians, looking for
the implications that this description has for a picture which
deals with subnuclear degrees of freedom (QMC model).
We are particularly interested in the EMC effect because it probes
directly the role of the nucleon structure in finite nuclei
\cite{ARNEODO}. In the present work we have studied the
reliability of a coherent description of in-medium nucleon
properties in terms of QHD and QMC models \cite{AGUIRRE2}.
Our approach considers the density dependence of the bag radius
and the parameters $B$ and $z_0$, motivated by recent researches
based on chiral symmetry restoration arguments \cite{XJ1}.
In ref. \cite{XJ1} it has been questioned the assumption of fixing
the MIT bag constant at its free-space value, adopted in the
standard QMC treatment. We have made a suitable selection of the
equilibrium conditions for the in-medium bags which agree with
the normal
QMC scheme at zero baryon density. We have obtained density
dependent bag parameters in a closed form, which is able to
describe the main features of the EMC effect. There exist another
models to calculate $B$ as a function of the density. For example
the models developed by Jin and Jennings \cite{XJ1,XJ2}
based on phenomenologycal considerations and the so-called
Global Colour Gluon Model (GCM) \cite{CAHILL}.
The parameter $B$ in the GCM is evaluated as a function of
the mean value of the scalar meson field,
taken into account the energy density for complete restoration
of the chiral symmetry inside a cavity.

In our work the bag parameters are obtained as functions of the
baryon density, taking their derivatives with respect to the mean
field value of the scalar meson as parameters. We have examined
the EMC effect in the framework of the dynamical rescaling
parameter \cite{AGUIRRE2} using the Local Density Approximation
(LDA). We have compared our results with those provided by Close
et al. \cite{CLOSE} and by Jin and Jennings \cite{XJ2}.

This work is organized as follows. In Section II a brief review of
the hadronic lagrangians and the QMC model is presented. The
stability conditions for bags in the nuclear medium are stablished.
The results for the bag radius and  parameters as functions
of the density are shown in Section III. The application to the EMC
effect is described in Section IV. Section V is devoted to the
conclusions of this work.

\section{FORMALISM}
\subsection*{\centerline{\small{\bf A. The quantum field theory of
hadrons}}}

Since the work of Walecka \cite{WALECKA83} theoretical approaches
to nuclear physics using hadron fields have been widely studied.
Nucleons $\psi$ and mesons $\sigma, \omega_{\mu}, \pi, b_{\mu}$
are the relevant degrees of freedom. In the simplest version
(the so-called QHD-I model) nucleons and only scalar ($\sigma$)
and vector ($\omega_{\mu}$) neutral mesons are used.
The nuclear matter saturation properties are adjusted with a
minimum number of free parameters, and the spin-orbit force is
implicitly included in the model. In the Mean Field Approximation
(MFA) \cite{SEROT} the mesons are considered as classical fields.
It has been argumented that this approach becomes increasingly
valid as the baryon density increases, and it has been succesfully
used to describe many aspects of nuclear matter \cite{SEROT} as
well as finite nuclei \cite{SEROT,BOGUTA}. Notwithstanding certain
undesirable features, such as a stiff equation of state, a rapidly
decreasing effective nucleon mass and increasing
effective meson masses have motivated alternative approaches.
Firstly, more involved schemes of approximation have been used.
The inclusion of vaccuum effects \cite{CHIN} and additional
contributions \cite{HOROWITZ} improve the results, but doing the
procedure excessively complicated.
Secondly, another forms of interaction have been proposed.
A polynomic self-interaction of the scalar field and consequently
additional free parameters have been included \cite{BOGUTA}.
In order to preserve the original renormalizability
this is a quartic polynomial.

Renormalizability implies that the quantum vacuum and all the
observables can be described in terms of hadronic degrees of
freedom exclusively. Since hadrons are bound states of quarks
and gluons this assumption must ultimately be invalid. This
situation must be evident where the short-distance effects are
dominants. Taking this into account non-renormalizable hadronic
models have been proposed and used to describe the nuclear
phenomenology in the MFA with convincing results
{\cite{ZM}}-{\cite{FELD}}. It is possible that the non-polynomic
character of these interactions simulates the effects of the
internal hadronic structure.

In the present work we have selected a set of four hadronic
models to evaluate the properties of symmetric nuclear matter.
For instance model M1 is the QHD-I model of Walecka
\cite{WALECKA83} and model M4 was proposed by Zimanyi and
Moskowszki \cite{ZM}. The models M2 and M3 \cite{FELD} present
an intermediate behavior between the previously mentioned models.

For all the cases the lagrangian density is given by
${\cal L}(x) = {\cal L}_0(x)+{\cal L}_{int}(x)$, where the free
fields sector is
\begin{eqnarray*}
{\cal L}_0(x)&=&\bar{\psi}(x)(i \not\!\partial - M_N)
\psi (x)+ \frac{1}{2} [\partial_\mu \sigma(x) \partial^\mu \sigma(x)
- m_\sigma^2 \sigma^2(x)] + \frac{1}{2} m_\omega^2 \omega_\mu(x)
\omega^\mu(x) \\
&&- \frac{1}{4} F^{\mu \nu}(x) F_{\mu \nu}(x)
\end{eqnarray*}
where $F^{\mu \nu}(x) =  \partial^\mu \omega^\nu (x) -
\partial^\nu \omega^\mu (x) $,
and the interaction term is
\begin{eqnarray*}
{\cal L}_{int}(x)&=& \bar{\psi}(x)[g_{\omega} \not \! \omega (x)
+ V(\sigma)] \psi(x)
\end{eqnarray*}

The masses of free nucleon, scalar and vector mesons have been
taken as $M_N=939$ MeV, $m_{\sigma}=550$ MeV and $m_{\omega}=783$
MeV, respectively. In Table I we present the explicit form of the
functional $V(\sigma)$ for each model and the corresponding coupling
constants ($g_{\sigma}$ and $g_{\omega}$) needed to adjust the nuclear matter
binding energy per nucleon, $E_b = 16$ MeV, and the
normal density of saturation, ${\rho} = 0.15$ fm$^{-3}$, in the MFA.
The isothermal compressibility
$\kappa = 9 {\rho}\,( \partial P_h/\partial \rho)_T$
is a usefull quantity in order to analyze the fitness of hadronic
models, since its value at the saturation nuclear density has been
well determined to range between $100-300$ MeV \cite{BLAIZOT}.
The isothermal compressibility evaluated in the MFA
is also shown in Table I \cite{FELD}.

In the MFA the fields $\sigma (x)$ and $\omega_{\mu} (x)$ are replaced
by their mean values which become constants in infinite homogeneus
nuclear matter, i.e.
$\sigma=\bar{\sigma}$ and $\omega_{\mu}=\bar{\omega} \delta_{\mu 0}$.
The Euler-Lagrange equations for nucleons and mesons in the MFA
are given by
\beq
(i {\not\!{\partial}} - M_N) {\psi}(x) =
(g_{\omega} {\bar{\omega}} {\gamma}_0 -
V({\bar{\sigma}})) {\psi}(x) ,
\eeq
\beq
m^2_{\sigma} {\bar{\sigma}} =
\left. \frac{d V}{d \sigma} \right |_{\bar{\sigma}}
\, \, \, {\rho}_s ,
\label{HADRONSELF}
\eeq
\beq
m^2_{\omega} {\bar{\omega}} = g_{\omega} {\rho} ,
\eeq
where ${\rho}_s = < {\bar{\psi}}(x) {\psi}(x) >$ is the scalar
density and ${\rho} = < {\bar{\psi}}(x) \gamma^0 {\psi}(x) >$ is
the baryon density. The nuclear matter density at zero temperature
is given by
\beq
{\rho}={{2 k^3_F}\over{3 {\pi}^2}} ,
\eeq
and
\beq
{\rho}_s= {{4}\over{(2 \pi )^3}} \int d^3{\vec{k}}
\Theta (k_F-|\vec{k}|)
{{M^*_N}\over{\sqrt{{M^*_N}^2+{\vec{k}}^2}}} ,
\eeq
where $k_F$ is the Fermi momentum. The effective nucleon mass
$M^*_N$ and the energy spectrum $\epsilon (k)$ are the followings
\beq
M^*_N = M_N - V({\bar{\sigma}}) ,
\label{HADRONMASS}
\eeq
and
\beq
\epsilon (k) = \sqrt{{M^*_N}^2+k^2} + g_{\omega} {\bar{\omega}} .
\eeq

The hadronic pressure for uniform nuclear matter,
$P_h = - {{1}\over{3}} T^{ii}$ (where $T^{ii}$ is the trace over
the spatial components of the energy-momentum tensor $T^{\mu \nu}$),
can be written as
\beq
P_h = \frac{2 M^{*3}_N}{\pi^2} \left[ \frac{k_F}{M^*_N}
\left(\frac {5 \epsilon_F}{24} - \frac{\epsilon_F^3}{12 M^{*2}_N}
\right) - \frac{M^*_N}{8} ln \left( \frac{\epsilon_F+k_F}{M^*_N}
\right) \right]  \,\, ,
\label{HADRONPRESS}
\eeq
where ${\epsilon}_F =  \sqrt{M^{* 2}_N+k_F^2}$.

Equation (\ref{HADRONSELF}) is a self-consistent definition for
the mean field value $\bar{\sigma}$, indeed from this equation
we see that the derivative  $d M^*_N/d\bar{\sigma}$ determines
the dynamics of the scalar field. As we have previously mentioned
the models considered have different degrees of accuracy in the
description of statistical as well as single particle nuclear
properties. If the validity of these results is assumed, one can
ask about the subnuclear dynamics compatible with such a behavior.
The QMC is a suitable model to describe the nucleon substructure
and, although it is a closed model by itself, we propose to treat
the bag interaction using the hadronic models. The meson fields are
the link between hadronic and effective quark models.

\subsection*{\centerline{\small{\bf B. The Quark-Meson
Coupling model}}}

We briefly recall some basic features of the QMC model
\cite{GUICHON,SAITO200}. If isospin symmetry breaking is neglected
the meson fields $\sigma (x)$ and $\omega_{\mu} (x)$ are
sufficient to describe the problem. The equation of
motion inside the bag for quarks of mass $m_q$ is given by
\beq
(i {\not\!{\partial}} - m_q) {\Psi}_q(x)
= [-g^q_{\sigma} \sigma (x) +
g^q_{\omega} {\not\!{\omega (x)}}] {\Psi}_q(x) ,
\eeq
where $g^q_{\sigma}$ and $g^q_{\omega}$ are the quark-meson coupling
constants associated with the $\sigma$ and $\omega_{\mu}$ fields,
respectively. Meson fields are	treated in the MFA  and
the continuity of these fields through the bag surface is
assumed \cite{SAITO51}.

The normalized ground-state quark wave function for a spherical bag
of radius $R$ is given by
\beq
{\Psi}_q({\vec{r}},t) = {\cal{N}} e^{-i {\epsilon}_q t/R}
\times \left( \begin{array}{c} {j_0(y r/R)} \\	i {\beta}_q
{\vec{\sigma}} \cdot $\^{r}$ j_1(y r/R)  \end{array} \right)
{{{\chi}_q}\over{\sqrt{4 \pi}}} ,
\eeq
where $r=|{\vec{r}}|$,
${\chi}_q$ is the quark spinor and the normalization constant is
\beq
{\cal{N}} = y / \sqrt{2 R^3 j_0^2(y)
[ {\Omega} ( {\Omega} - 1) + R m^*_q/2]} .
\eeq

We have introduced the effective quark mass
$m^*_q = m_q -g^q_{\sigma} {\bar{\sigma}}$ and the
effective energy eigenvalue
${\epsilon}_q = \Omega/R+ g^q_{\omega} {\bar{\omega}}$,
where ${\Omega} = \sqrt{y^2 + (R m^*_q)^2}$. Since $m_q$ is a free
parameter of the QMC model, in this work we have used the value
$m_q=10$ MeV. The $y$ variable is
fixed by the boundary condition at the bag surface
$j_0(y) = {\beta}_q j_1(y)$, which implies a zero normal flow of
the quark current through the bag surface \cite{CHODOS,THOMAS1}.
Here ${\beta}_q = \sqrt{({\Omega} - R m^*_q)/({\Omega} + Rm^*_q)}$.

The bag energy is given by
\beq
E_b = {{3 {\Omega} - z_0}\over{R}} + {{4}\over{3}} \pi B R^3 \,\, ,
\label{BAGENERGY}
\eeq
where $B$ is the energy per unit of volume and $z_0$ takes into
account the zero point energy of the bag. The nucleon mass is
defined by including the correction due to the spurious center
of mass motion \cite{CMC}
\beq
M^*_b = \sqrt{E_b^2 - 3 {(y/R)}^2} .
\label{BAGMASS}
\eeq
Different notations for the nucleon mass entering in QHD ($M^*_N$)
and the nucleon mass generated by the bag model ($M^*_b$) are used
in this work. Both will be appropriately related below.

For the internal pressure inside the bag, defined as
$P_b=- \partial E_b /
\partial V$, we have
\begin{eqnarray}
P_b  &=&-\frac{E_b}{4 \pi R^2  M^*_b}\left(-\frac{E_b}{R}+
\frac{16}{3}
\pi R^2 B+ \frac{3 m^{* 2}_q}{\Omega} \right)-
\frac{3 y^2}{4 \pi R^5 M^*_b} .
\label{BAGPRESS}
\end{eqnarray}

It is usual to determine the bag parameters at zero baryon density
to reproduce the experimental nucleon mass $M_b = 939$ MeV.
Simultaneously it is required the equilibrium condition for the bag
$d M_b( \sigma )/d R = 0$, evaluated at $\sigma=\bar{\sigma}$.
By using the bag radius at zero baryon density $R_0$, and the quark
mass $m_q$ as free parameters, one can obtain the values shown in
reference \cite{SAITOHEP}. In this reference the MIT bag parameters
$B$ and $z_0$ are constants, although it is expected a density
dependence for the bag radius and the quark effective mass.
In fact, it was found \cite{SAITOHEP} a relative change of $1 \%$
in the radius at the saturation density (${\rho}=0.15$ fm$^{-3}$)
as compared with its value at zero baryon density.

\subsection*{\centerline{\small{\bf C. Equilibrium conditions for
bags in nuclear medium}}}

If QMC and QHD produce coherent descriptions, the following
condition must be fulfilled
\beq
M^*_N( \sigma ) = M^*_b ( \sigma ),
\label{IGUALMASS}
\eeq
together with $g_{\sigma}=3 g^q_{\sigma}$,
$g_{\omega}=3 g^q_{\omega}$, \cite{SAITO200}.

The stability in the nuclear medium with respect to volume changes
is impossed by
\beq
P_b( \sigma )=P_h( \sigma ) ,
\label{IGUALPRESS}
\eeq
where $P_b( \sigma )$ is the internal pressure generated by the
quark dynamics Eq. (\ref{BAGPRESS}) and $P_h( \sigma )$ is the
external hadronic pressure Eq. (\ref{HADRONPRESS}).

Equation (\ref{IGUALPRESS}) is a statistical equilibrium condition
on the bag surface which ensures a direct relation between nuclear
matter bulk properties and the stability of the confining volume.
In the standard QMC model the equilibrium condition for the bags
is $dM^*/dR = 0$. This condition implies the zero bag pressure in
all the cases (free and interacting bags) and provides
a density dependent bag radius. However, the qualitative behavior
obtained for $R$ \cite{SAITOHEP} is not sufficient to
fit the enhancement of the confinement volume expected in the
framework of the dynamical rescaling parameter \cite{CLOSE} for
the EMC effect.

The functional relations Eq. (\ref{IGUALMASS}) and
Eq. (\ref{IGUALPRESS}) can be used to get the density variation
of the bag parameters. They can be expanded in power series and
by equating coefficient to coefficient additional relations can
be obtained. Since in our approach only the linear contributions
have been retained, in addition to Eq. (\ref{IGUALMASS})
and Eq. (\ref{IGUALPRESS}) we have the following conditions
\beq
\left. {{d M^*_b(\sigma)}\over{d \sigma}} \right |_{\bar{\sigma}} = -
\left. {{d V(\sigma)}\over{d \sigma}}\right |_{\bar{\sigma}} \,\, ,
\label{IGUALDMASS}
\eeq
\beq
\left. {{dP_b(\sigma)}\over{d \sigma}}\right |_{\bar{\sigma}} =
\left. {{dP_h(\sigma)}\over{d \sigma}}\right |_{\bar{\sigma}} \,\,  ,
\label{IGUALDPRESS}
\eeq
for every ${\bar{\sigma}}$ at a given density.

In order to clarify this work we present a detailed form
for the expressions used in our calculations.
By using equations (\ref{BAGENERGY}) and (\ref{BAGMASS})
one can evaluate the following derivatives
\begin{eqnarray}
\frac{d M^*_b}{d \bar{\sigma}} =
\left( \frac{\partial M^*_b}{\partial m^*_q}
\right)_{y,R,z_0,B} \frac{dm^*_q}{d \bar{\sigma}} +
\left( \frac{\partial M^*_b}{\partial y}\right)_{m^*_q,R,z_0,B}
\frac{d y}{d \bar{\sigma}}+
\left( \frac{\partial M^*_b}{\partial R}\right)_{y,m^*_q,z_0,B}
\frac{dR}{d \bar{\sigma}}   \nonumber \\
+ \left( \frac{\partial M^*_b}{\partial z_0}\right)_{y,m^*_q,R,B}
\frac{dz_0}{d \bar{\sigma}} +
\left( \frac{\partial M^*_b}{\partial B}\right)_{y,m^*_q,R,z_0}
\frac{dB}{d \bar{\sigma}} ,
\end{eqnarray}
where
\begin{eqnarray}
\left( \frac{\partial M^*_b}{\partial m^*_q}\right)_{y,R,z_0,B}&=&
\frac{3 E_b R m^*_q}{\Omega  M^*_b},  \\
\left( \frac{\partial M^*_b}{\partial y}\right)_{m^*_q,R,z_0,B}&=&
\frac{3 y}{R M^*_b}\left(\frac{E_b}{\Omega}-\frac{1}{R} \right) ,\\
\left( \frac{\partial M^*_b}{\partial R}\right)_{m^*_q,y,z_0,B}&=&
\frac{E_b}{M^*_b}\left(-\frac{E_b}{R}+\frac{16 \pi}{3} R^2 B+
\frac{3 m^{* 2}_q}{\Omega} \right)+\frac{3y^2}{M^*_bR^3} ,\\
\left( \frac{\partial M^*_b}{\partial z_0}\right)_{m^*_q,y,R,B}&=&
-\frac{E_b}{R M^*_b} ,\\
\left( \frac{\partial M^*_b}{\partial B}\right)_{m^*_q,y,R,z_0}&=&
\frac{4}{3} \frac{\pi E_b R^3}{M^*_b} .
\end{eqnarray}

Since the $\sigma$-dependence is contained in the bag parameters
$R$, $B$ and $z_0$ our hypotesis is equivalent to assume that the
derivatives $\lambda=dB/d\bar{\sigma}$ and
$\mu=dz_0/d\bar{\sigma}$ are reduced to constant values.

\section{DENSITY DEPENDENCE OF THE BAG PARAMETERS}

\subsection*{\centerline{\small{\bf A. Determination of
$(\lambda,\mu)$}}}

To search for appropriate values of $\lambda$ and $\mu$
we have explored the $(\lambda, \mu)$ plane at zero
baryon density. Fixing the zero density bag radius $R_0$
we have found a linear relation $\lambda = \lambda(\mu)$
for all the models, as it is shown in Fig.1(a). The slope
of these straight lines $\lambda ( \mu )$ does not
depend on the specific hadronic model used.
The linear dependence is preserved for higher densities,
but now the slope depends on the effective hadronic model
(see Fig.1(b)).

Futhermore we have evaluated $R, B$ and $z_0$ at zero baryon
density as a function of $\lambda$ for several values of $\mu$
and using the model M4. It was found a drastic change in the
regime of variation of the quantities considered when $\mu$
goes from negative to positive values as can be seen in Figs.
2(a), 2(b) and 2(c). Indeed we have found two regions: in the
first one, where $\mu < 0$ (region I), the  parameters $B$ and
$z_0$ are monotonous decreasing functions of $\lambda$ while
$R$ is monotonous increasing. In the other region, where
$\mu > 0$ (region II), the behavior of each quantity has been
reversed as compared with region I.
At the limit of these two regions a discontinuity is found.
The location of this discontinuity
point is only weakly dependent on $\mu$ and is given
approximately by $\lambda \simeq 0$.
It must be pointed out that all the hadronic models produce a
similar behavior.

In addition we have examined the bag radius $R$
at the saturation density as a function of $\lambda$
by fixing its value at zero baryon density  $R_0=0.6$ fm.
For the hadronic models considered here we have found that
$R/R_0>1$ only for a restricted range of $\mu$ belonging to
region II. However, the accurate situation of this interval
changes from one model to another. For the following discussion
we have taken two sets of values; set I ($\lambda=-5.28$ fm$^{-3}$,
$\mu=-0.50$ fm) and set II ($\lambda=0, \mu=1.6$ fm)
which places the model M4 in the regions I and II, respectively.

\subsection*{\centerline{\small{\bf B. The bag parameters and
radius}}}

After the hadron coupling constants have been adjusted to reproduce
the saturation conditions for nuclear matter, Eq. (\ref{HADRONSELF})
can be used to obtain the self-consistent solution $\bar{\sigma}$
for each baryon density. Equations (\ref{HADRONMASS}),
(\ref{BAGENERGY}), (\ref{BAGMASS}), (\ref{IGUALMASS}) and
(\ref{IGUALDMASS})
can be used together with Eqs. (\ref{HADRONPRESS}),
(\ref{BAGPRESS}), (\ref{IGUALPRESS}) and
(\ref{IGUALDPRESS}) to determine $R$, $B$ and $z_0$
for fixed values of the parameters $\lambda$ and $\mu$.

The bag radius as a function of the density is shown in Fig.3.
It can be seen that set I gives an asymptotical constant bag radius
for every model, these constant values do not depend on the details
of the
interaction and they are diminished as compared with its vacuum value.
The set II provides a model dependent radius at high densities, in
this case models M1 and M2 predict a breaking-down of the bag
picture. Models M3 and M4 have a stable behavior for all the
densities considered even when set II is used.

In Fig.4 and Fig.5 we present the density dependence of $B^{1/4}$ and
$z_0$, respectively. Models M1 and M2 exhibit an opposite
behavior for $B^{1/4}$ when set I or set II are used.

With respect to the behavior of model M1 it can be seen that
for set I, $B$ takes negative values as the baryon density is
sufficiently increased, thus the bag bulk energy must decrease
with increasing volume. The increment of the kinetic energy
compensates this fact, giving a slowly decreasing
total bag energy and a stable bag radius (see Fig.3).
On the other hand, $B$ grows drastically at high densities
for set II, a small volume increment gives rise to a large
increment in the bulk energy. Therefore in order to get a slowly
decreasing $M^*_b$ the bag radius $R$ must decrease at the same
rate as $B^{1/3}$ grows. When $R$ approaches to zero a subtle
cancellation among the quark kinetic energy, the zero point
energy parameter and the center of mass correction takes place.
The raising of $z_0$ (see Fig.5) is not sufficient to reach the
dynamical equilibrium and hence the system reduces its volume as
far as possible.

The steeper behavior of models M1 and M2 as compared with models
M3 and M4 is due to the fact that the first mentioned models give
a stiffer equation of state and a faster decrease for the effective
nucleon mass. Furthermore models M1 and M2 have a more noticeable
dependence on the set of parameters used.

Our results can be compared with those obtained by Jin and Jennings
\cite{XJ1}.
In ref. \cite{XJ1} the density dependence of the bag constant
has been modeled in two different forms: the so-called
Direct Coupling Model (DCM) and the Scaling Model (SM).
The bag constant is parametrized as follows
\beq
\frac{B}{B_0} = {\left[ 1 - 4 \frac{g^B_{\sigma}
{\bar{\sigma}}}{\delta M_N} \right]}^{\delta}	\,\,\, ,
\eeq
for the DCM and
\beq
\frac{B}{B_0} = {\left[ \frac{M^*_N}{M_N}
\right]}^{k},
\eeq
for the SM.
Where $g_{\sigma}^B$, $\delta$ and $k$ are positive
parameters and $B_0$ is the bag parameter at zero density.

The DCM is partially motivated from considering a non-topologycal
soliton model for the nucleon where the scalar soliton field
provides the quarks confinement. Within this framework
a monotonous density dependence of $B$ and $R$ is found.

In Table II we show a comparison between our results by using model
M4 with the set II and the ones of
ref. \cite{XJ1} with the DCM. We have selected model M4 due
to its smooth density dependence and the set II of parameters
because it produce an increasing bag radius, as we have
previously discussed.

The last row of this table shows our calculations,
in the remaining rows the results of ref. \cite{XJ1} are
presented. It can be seen that our results for $B/B_0$
and $R/R_0$ are similar to those corresponding to $g_{\sigma}^q=2$
and $\delta = 12$ in the DCM, but the compressibility is
appreciably lower for our calculations.

The parameter $B$ as a function of ${\bar{\sigma}}$ for
the DCM, SM and M4 models is shown in Fig.6.
The curve corresponding to M4 (solid line)
has been fitted with a quadratic polynomial
$B=A_0+A_1 {\bar{\sigma}}+A_2 {\bar{\sigma}}^2$ (short-dashed line).
The best fit has been obtained for $A_0=184,723$ MeV,
$A_1=-186.42$ MeV and $A_2=112.37$ MeV. These values correspond to
$g_{\sigma}^q=4.8$ and $\delta = 20.5$ (long-dashed line) for the
DCM in the same degree of approximation.
The dotted-line curve shows the results obtained using the SM with
$k=3.16$.
Our results are in good agreement with the SM model and
they coincide with the DCM model only
for small values of ${\bar{\sigma}}$.

\section{APPLICATION TO THE EMC EFFECT}

\subsection*{\centerline{\small{\bf A.
The EMC effect in the dynamical rescaling mechanism}}}

The EMC effect accounts for the distortion of the
nucleon structure functions due to the presence of other hadrons
\cite{ARNEODO,CLOSE},
as it is well known from deep inelastic scattering experiences
the structure function per nucleon
differs in the case of bound or free nucleons.
In the range $0.2 < x < 0.6$, where $x$ is the Bjorken variable,
it is expected that $F_2^A(x,Q^2)/F_2^N(x,Q^2) < 1$. Here
$F_2^A(x,Q^2)$ is the structure function for a nucleon in a nucleus
of atomic number $A$, $F_2^N(x,Q^2)$ is the corresponding
structure function for free nucleons and $Q^2$ is the probing
momentum \cite{ARNEODO,CLOSE}.

Explanation of the EMC effect has given rise to a lot of
theoretical
work, the formalisms used ranges from traditional nuclear
descriptions in terms of pion exchange \cite{EMC-pi} or binding-
energy shifts \cite{EMC-binding,EMC-GarciaC} to QCD inspired
models such as
dynamical rescaling \cite{CLOSE}, multiquark clustering
\cite{EMC-clustering} and deconfinement in nuclei
\cite{EMC-deconf}. Some of them provides a good fit to the
$x$-dependence of the phenomenon in spite of the quite different
underlying assumptions. In consequence there not exist an
unambiguous discrimination of the origin of the EMC effect.

An intermediate treatment was given in \cite{EMC-QMC,EMC-QMCb}.
In these works
meson fields and explicit quark dynamics are included in the
evaluation of the twist-2 valence quark distribution. By using the
LDA for finite nuclei a good semiquantitative description of the
$^{56}$Fe data is obtained.

In order to test our previous results in this section
we try to describe the EMC effect.
Our approach is in the trend of \cite{EMC-QMC,EMC-QMCb} in
the sense that a
mixed meson-exchange formalism and MIT bag model are linked, however
the effect is focused in the dynamical rescaling framework
\cite{CLOSE}.
Thus our work is closer to  \cite{XJ2}, furthermore since we
have included a non-linear meson-nucleon interaction from the outset
we do not need to complete the QMC model as in \cite{EMC-QMCb}.
We must remark that the density dependence of the confinement scale
is determined by the hadronic interaction, consequently
in our approach the QCD
inspired dynamical rescaling and the meson exchange are not
excluding mechanisms.

We have not considered shadowing and Fermi motion and
we have restricted to the region $0.2 < x < 0.6$,
where the valence quark role is dominant.

In the dynamical rescaling framework an increase
of $15 \%$ is expected for the bag radius at the saturation density
to explain the experimental observations for $^{56}Fe$ \cite{CLOSE}.

This dynamical rescaling is based on the fact that the free nucleon
confinement size is smaller than that corresponding to a
bound nucleon in a nucleus. The rescaling relation can be written as
\beq
F^A_2(x,Q^2) = F^N_2(x, \zeta_A Q^2) .
\eeq
Here $\zeta_A (Q^2)$ is the rescaling parameter
associated with the ratio of the quark
confinement size in the nucleus $A$ and the free radius
\beq
\zeta_A(Q^2) = {\left( {{{\bar{R}}_A}\over{R_0}}
\right)}^{\eta ( \mu ,
Q^2)} ,
\eeq
where
\beq
\eta ( \mu , Q^2) = 2 {{ln(Q^2/{\Lambda}^2_{QCD})}\over
{ln({\mu}^2/{\Lambda}^2_{QCD})}} .
\eeq

Here $\bar{R}_A$ is the average radius of the nucleon in a nucleus A,
and $R_0$ is the corresponding radius for a free nucleon.
The mass scale parameter ${\Lambda}_{QCD}$ is related to the
$Q^2$-dependence of the QCD coupling constant and the scale
parameter $\mu$
remains unspecified in perturbative QCD and can be interpreted as
a low momentum cutoff for gluon radiation. Therefore for
$Q^2 = {\mu}^2$ the nucleon appears as a bound state of three
valence quarks. In order to compare results
we have taken ${\Lambda}_{QCD} = 0.25$ GeV,
${\mu}^2 = 0.66$ GeV$^2$ and $Q^2=20$ GeV$^2$ \cite{XJ2}.

\subsection*{\centerline{\small{\bf B. Results for the EMC effect}}}

From the density dependence of $R$, $B$ and $z_0$ obtained for
nuclear matter it is possible to evaluate the averages
of these quantities for finite nuclei in the
LDA. In this approach the nucleon
average radius inside a finite nucleus of
atomic number $A$, is obtained by using
\beq
{\bar{R}}_A = {\cal{M}} \int d^3 {\vec{r}} R[{\rho}_A(r)]
{\rho}_A(r) ,
\eeq
where ${\rho}_A(r)$ is the local density distribution for
the nucleus considered,
$R[{\rho}_A(r)]$ denotes the nuclear matter bag radius at the
density ${\rho}_A(r)$
and ${\cal{M}}^{-1} = \int d^3 {\vec{r}} {\rho}_A(r)$.
The bag parameters $B$ and $z_0$ can be averaged in a similar way
\beq
{\bar{B}}_A = {\cal{M}} \int d^3 {\vec{r}} B[{\rho}_A(r)]
{\rho}_A(r) ,
\eeq
and
\beq
{\bar{z_0}}_A = {\cal{M}} \int d^3 {\vec{r}} z_0[{\rho}_A(r)]
{\rho}_A(r) ,
\eeq
where $B[{\rho}_A(r)]$ and $z_0[{\rho}_A(r)]$
denote the bag parameters $B$ and $z_0$,
at the density ${\rho}_A(r)$.
For the nucleon density distribution in a finite nucleus
we have used the standard Woods-Saxon
expression
\beq
{\rho}_A(r) = {{{\rho}^0_A}\over{1+exp[(r-R_A)/a_A]}} .
\eeq
The parameters ${\rho}^0_A$, $R_A$ and $a_A$ are listed in
reference \cite{DENSI} and they are fixed to fit nuclear shapes.
Our results for the rescaling parameter are presented in Table III,
where model M4 and set II
have been selected. There we have compared them
with those of \cite{XJ2} and \cite{CLOSE}.
Our results are in good agreement with the ones presented in
ref. \cite{CLOSE}, the best fit is obtained for $^{197}Au$ with a
discrepancy of $1 \%$, meanwhile for $^{118}Sn$ and $^{208}Pb$
the disagreement is of $4 \%$ approximately.
The fit is worse for light nuclei, this is due to the fact that the
LDA is more adecuate for heavy nuclei where the surface effects are
reduced and the Woods-Saxon density fits better.
For the cases considered in Table III the proton-neutron
asymmetry increases with the baryon number $A$ and this effect
has not been taken into account in our calculations.
Thus the best fit obtained corresponds to a
sufficient heavy nuclei which
compensates the proton-neutron asymmetry.
This treatment can be improved starting from asymmetric nuclear
matter.

Models M1-M3 do not reproduce the qualitative
behavior of the rescaling parameter, even when the set II of
parameters is used.

\section{CONCLUSIONS}

We have studied the coherence of the QHD and QMC descriptions
by using the equilibrium conditions for the MIT bag in
nuclear matter, Eqs. (\ref{IGUALMASS}) and (\ref{IGUALPRESS}).

The density dependence of the bag parameters
and the bag radius have been evaluated by using two dynamical
quantities,
i.e. the derivatives $dB/d \bar{\sigma}$ and $dz_0/d \bar{\sigma}$,
as parameters and we have explored their possible variation range.
We have found two different dynamical regimes for these parameters.
Only the region with $\mu >0$ can produce results for $B$ which
agree with those expected from chiral restoration arguments
\cite{BROWN,XJ1}. A linear relation between $\lambda$ and $\mu$
has been obtained at fixed bag radius  for all the models and the
considered densities.

We have chosen the Zimanyi-Moszkowski model to study the EMC effect
because it is the more adequate to describe bulk properties of
nuclear matter in the MFA \cite{ZM,SH}.
For finite nuclei, in the LDA and by using the
dynamical rescaling hypothesis we have obtained a good description
of the EMC effect in a complete self-consistent calculation.

We must remark that in our approach we have not introduced
phenomenologycal parameters and the considered ones can be
dynamically derived, furthermore dynamical rescaling and meson
exchange effects are not excluding mechanisms, instead the rescaling
parameter is derived from the meson background properties.

An appropriated treatment of the N-Z asymmetry	and the
inclusion of the $\rho$-meson in the models of the field theory of
hadrons could improve the present results.
Inclusion of thermal effects as well as consideration of the N-Z
asymmetry in the treatment of the EMC effect will be reported
in future works.

\section{ACKNOWLEDGMENTS}
This work was partially supported by the
University of La Plata, Argentina.

\newpage

\begin{table}
\caption{Nucleon-scalar meson interaction terms,
coupling constants and isothermal
compressibility $\kappa$, for several effective hadronic models
used in this work.}

\vspace{0.4cm}

\begin{tabular}{ccccc}
$Model$  &$V( \sigma ) $
& $g_{\sigma}$	& $g_{\omega}$ & $\kappa$ \\ \hline
	 &
&		&	       & [MeV] \\ \hline
$  M1  $  &$  g_{\sigma} {\sigma}$
 &    11.04	 &   13.74	&  554	\\
$  M2  $  &$M_N g_{\sigma}tanh(g_{\sigma} \sigma /M_N)$
 &     9.15	 &   10.52	&  410	\\
$  M3  $  &$M_N [1-exp(g_{\sigma} \sigma /M_N)]$
 &     8.34	 &    8.19	&  267	\\
$  M4  $  &$g_{\sigma}{\sigma}/(1+ g_{\sigma}\sigma /M_N)
$&     7.84	 &    6.67	&  224	\\
\end{tabular}
\end{table}

\newpage

\vfill
\eject
\begin{table}

\caption{Comparison between the DCM of ref. [21]
and our results. The last row corresponds to our calculations
with model M4 and set II, the remaining rows display the DCM
results corresponding to the model parameters indicated.}

\vspace{0.4cm}

\begin{tabular}{cccccccc}
$\delta$&$g_{\sigma}^q$&$g_{\sigma}^B$&$g_\omega$&$M_N^*/M_N$&$
\kappa$&$B/B_0$&$R/R_0$  \\
	&	       &	      & 	 &	     &
 [MeV]	&	&	  \\ \hline
   3.6	&      2       &    6.30      &   10.04  &  0.70     &
 431	&  0.36 &  1.27   \\
   4	&	       &    6.13      &    9.65  &  0.72     &
 398	&  0.39 &  1.25   \\
   8	&	       &    5.65      &    8.54  &  0.77     &
 336	&  0.48 &  1.18   \\
  12	&	       &    5.53      &    8.29  &  0.76     &
 324	&  0.50 &  1.17   \\ \hline
 set II &    2.61      &	      &    6.67  &  0.78     &
 224	&  0.54 &  1.12   \\

\end{tabular}

\end{table}

\begin{table}

\caption{The rescaling parameter $\zeta_A (Q^2)$ evaluated for
different
nucleus at $Q^2 = 20$ GeV$^2$. The superscripts $a$ and $b$ indicate
the results obtained by Jin and Jennings
[21] using DCM (with
$\delta =4$) and SM, respectively. Superscript $c$ indicates
our results using model M4 and the set II of parameters. The column
labeled $d$ corresponds to the reference of Close et al.
[24].
In all the cases $R_0=0.6$ fm.
In the last two columns the
averaged parameters $\bar{B}^{1/4}_A$ and $\bar{z}_{0A}$ calculated
in our scheme are shown.}

\vspace{0.4cm}

\begin{tabular}{cccccccc}
  $Nucleus$&${\zeta}_A(Q^2)^a$&${\zeta}_A(Q^2)^d$&${\zeta}_A(Q^2)^c$
&${\zeta}_A(Q^2)^d$&$\bar{B}^{1/4}_A$&$\bar{z}_{0A}$ \\
	   &		      & 		 &
&		   &   [MeV]   &	 \\ \hline
  $^{12}C$ &   1.69	      &    1.70 	 &    1.89
&    1.60	   &  169.94   & 1.949	 \\
  $^{20}Ne$&   1.69	      &    1.70 	 &    1.93
&    1.60	   &  169.22   & 1.944	 \\
  $^{27}Al$&   1.88	      &    1.89 	 &    2.06
&    1.89	   &  167.23   & 1.932	 \\
  $^{56}Fe$&   2.00	      &    2.02 	 &    2.16
&    2.02	   &  165.72   & 1.922	 \\
  $^{63}Cu$&   1.95	      &    1.96 	 &    2.17
&    2.02	   &  165.61   & 1.921	 \\
 $^{107}Ag$&   2.04	      &    2.07 	 &    2.36
&    2.17	   &  162.84   & 1.903	 \\
 $^{118}Sn$&   2.06	      &    2.09 	 &    2.34
&    2.24	   &  163.16   & 1.905	 \\
 $^{197}Au$&   2.24	      &    2.27 	 &    2.44
&    2.46	   &  161.79   & 1.896	 \\
 $^{208}Pb$&   2.16	      &    2.18 	 &    2.45
&    2.37	   &  161.73   & 1.896	 \\

\end{tabular}

\end{table}


\newpage

\vfill
\eject
\begin{figure}
\caption{
The parameter
$\lambda$ as a function of $\mu$ for fixed value of the bag radius
$R_1=0.6$ fm, $R_2=0.8$ fm and $R_3=1.0$ fm at zero density
(Fig.1(a)) and at the nuclear matter
saturation density (Fig.1(b)). In Fig.1(a) solid lines are used
for all the radii, in Fig.1(b) dot-dashed, dashed and solid lines
correspond to $R_1, R_2$ and $R_3$, respectively. In Fig.1(b)
for a fixed radius
and at low values of $\mu$, the values of $\lambda$ belonging to
models M1, M2, M3 and M4 are successively increased.
}

\end{figure}

\begin{figure}
\caption{
MIT bag parameters as a function of $\lambda$ for different
values of $\mu= -10$, $-5$, $-1$, $-0.1$, $0$, $0.1$, $1$, $5$
and $10$ fm  at zero
baryon density for the hadronic model M4. The curves
on the left (right) side of each figure corresponds to $\mu <0$
($\mu >0$), as $\mu$ increases the curves move in the way
indicated by the arrows.
The parameters $B$, $R$ and $z_0$ are shown in figures 2(a), 2(b)
and 2(c), respectively.
}

\end{figure}

\begin{figure}
\caption{
MIT bag radius $R$
as a function of the relative baryon density ${\rho}/{\rho}_0$,
for the considered hadronic models,
$\rho_0$ is the nuclear matter saturation density.
Full and dashed lines correspond to set I and set
II of parameters, respectively.
The quark mass at zero baryon density has been
taken as $m_q=10$ MeV.
}

\end{figure}

\begin{figure}
\caption{$B^{1/4}$
as a function of the relative baryon density ${\rho}/{\rho}_0$.
The notation and the line conventions are the same as in Figure
3.
}

\end{figure}

\begin{figure}
\caption{
$z_0$
as a function of the relative baryon density ${\rho}/{\rho}_0$.
The notation and the line conventions are the same as in Figure
3.
}

\end{figure}

\begin{figure}
\caption{
$B^{1/4}$ as a function of ${\bar{\sigma}}$. Solid, long-dashed,
short-dashed and dotted lines are corresponding to
M4 with set II, DCM, the polynomial regresion, and SM, respectively.
The values of the parameters used are indicated in the text.
}
\end{figure}

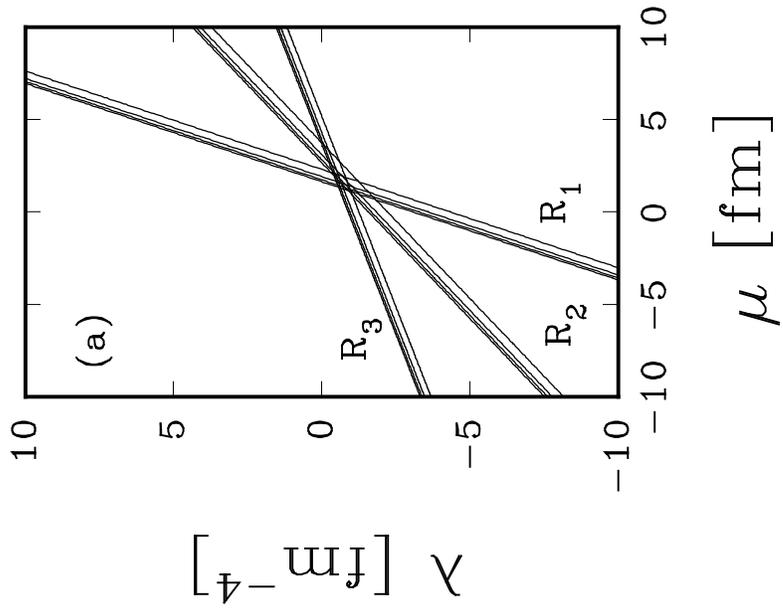
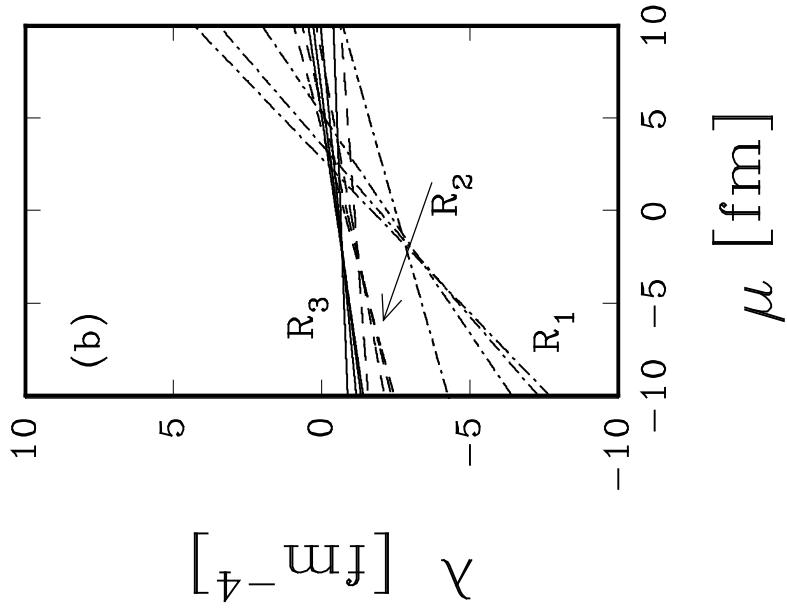
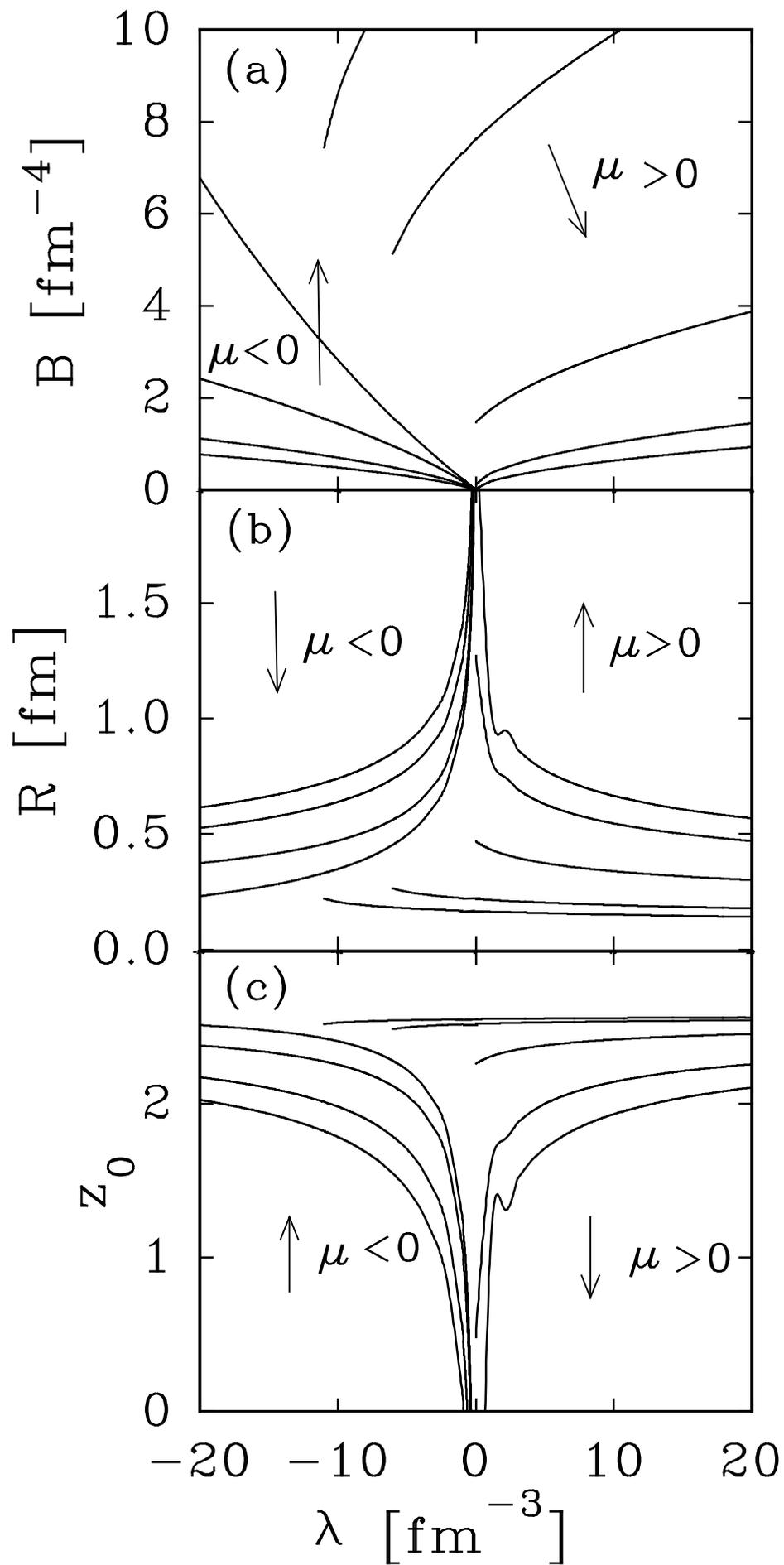
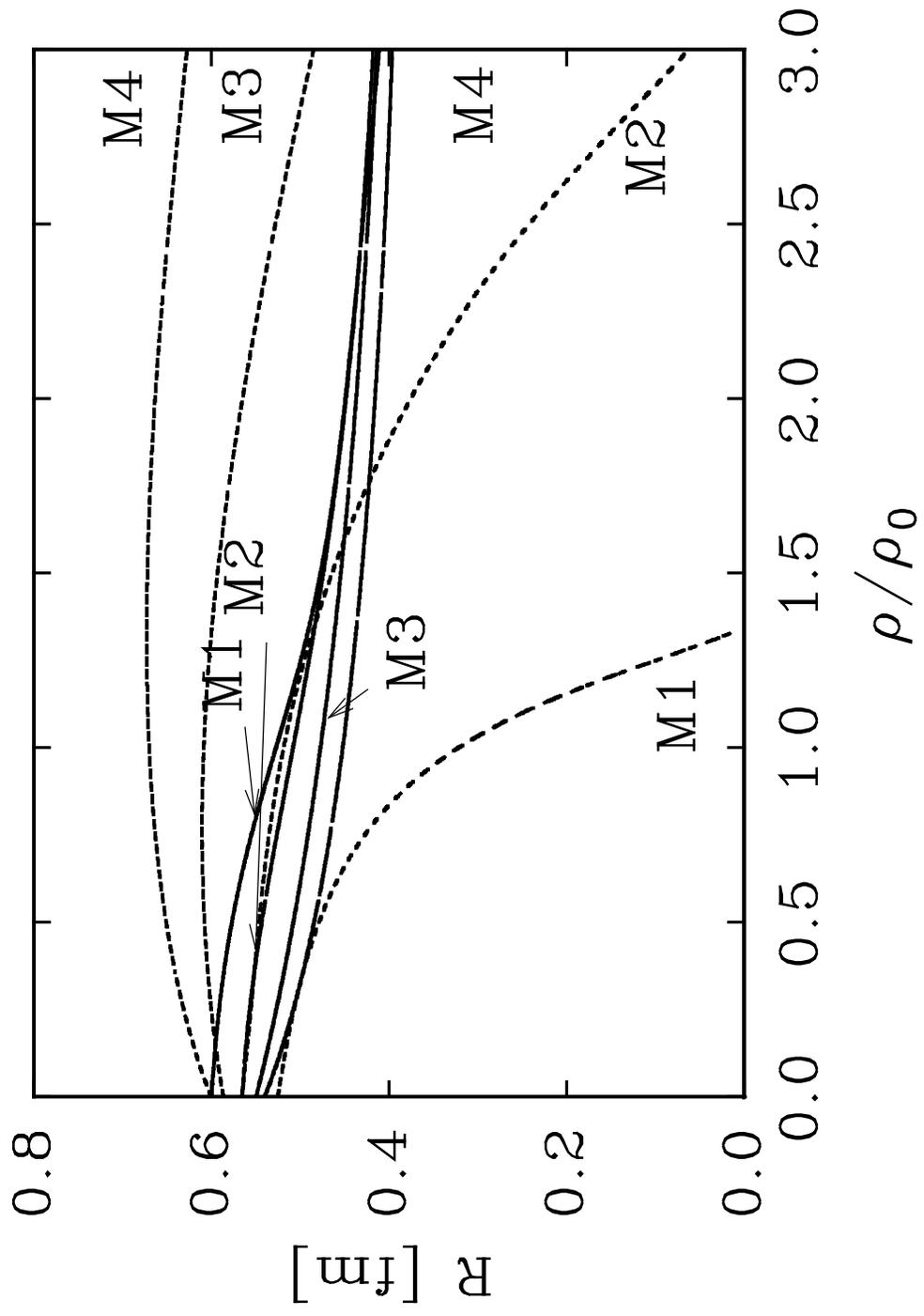
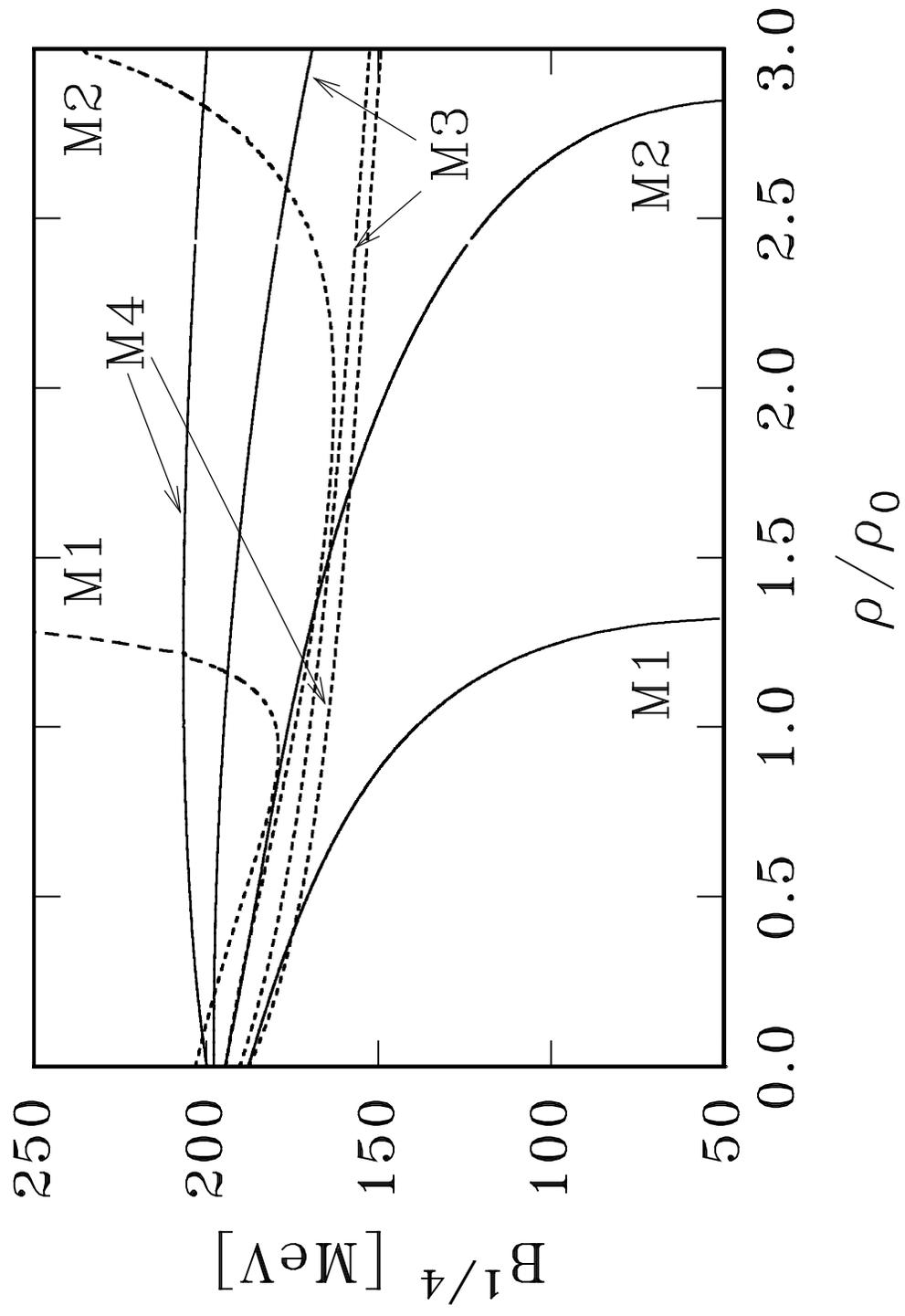
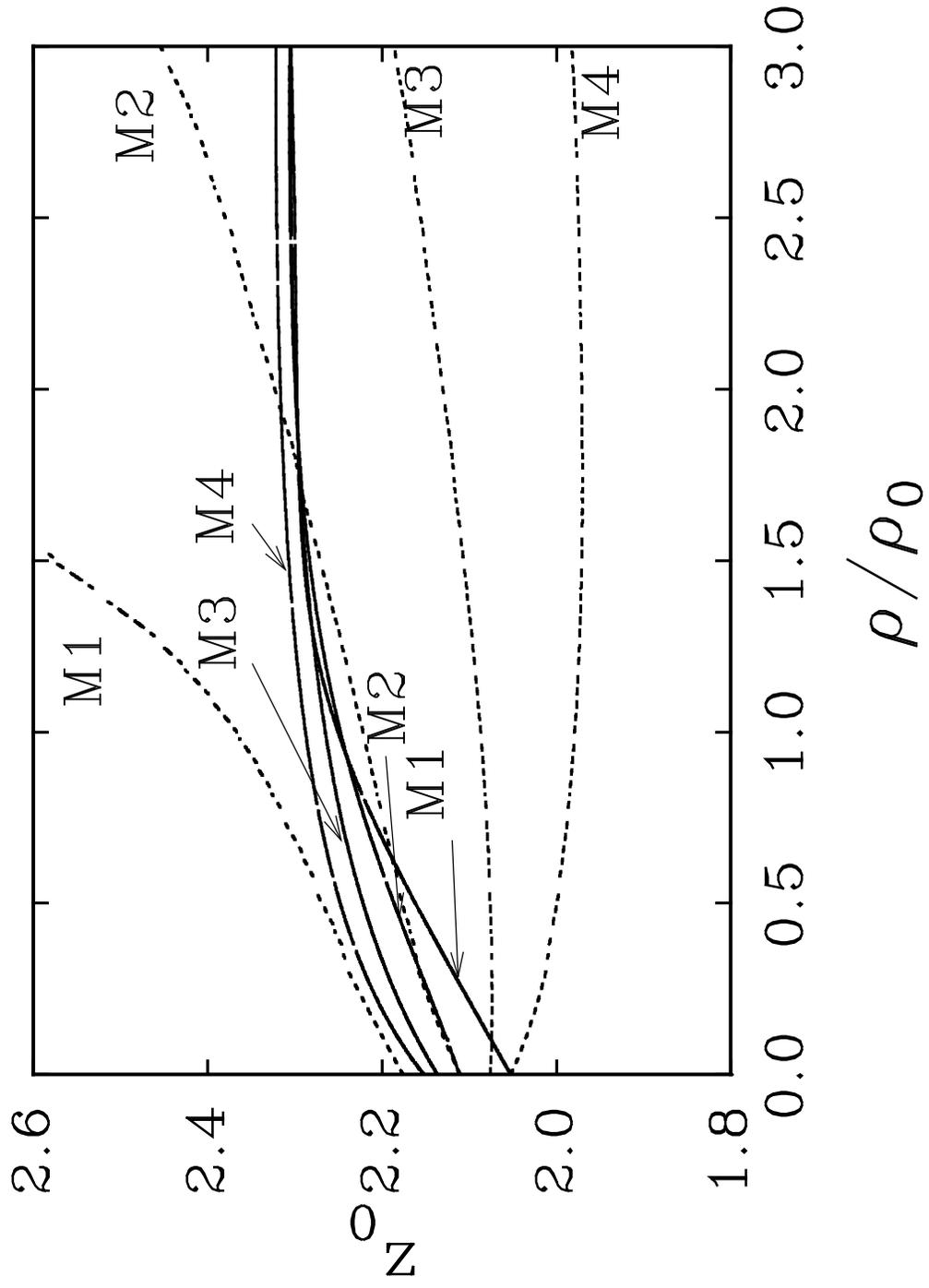
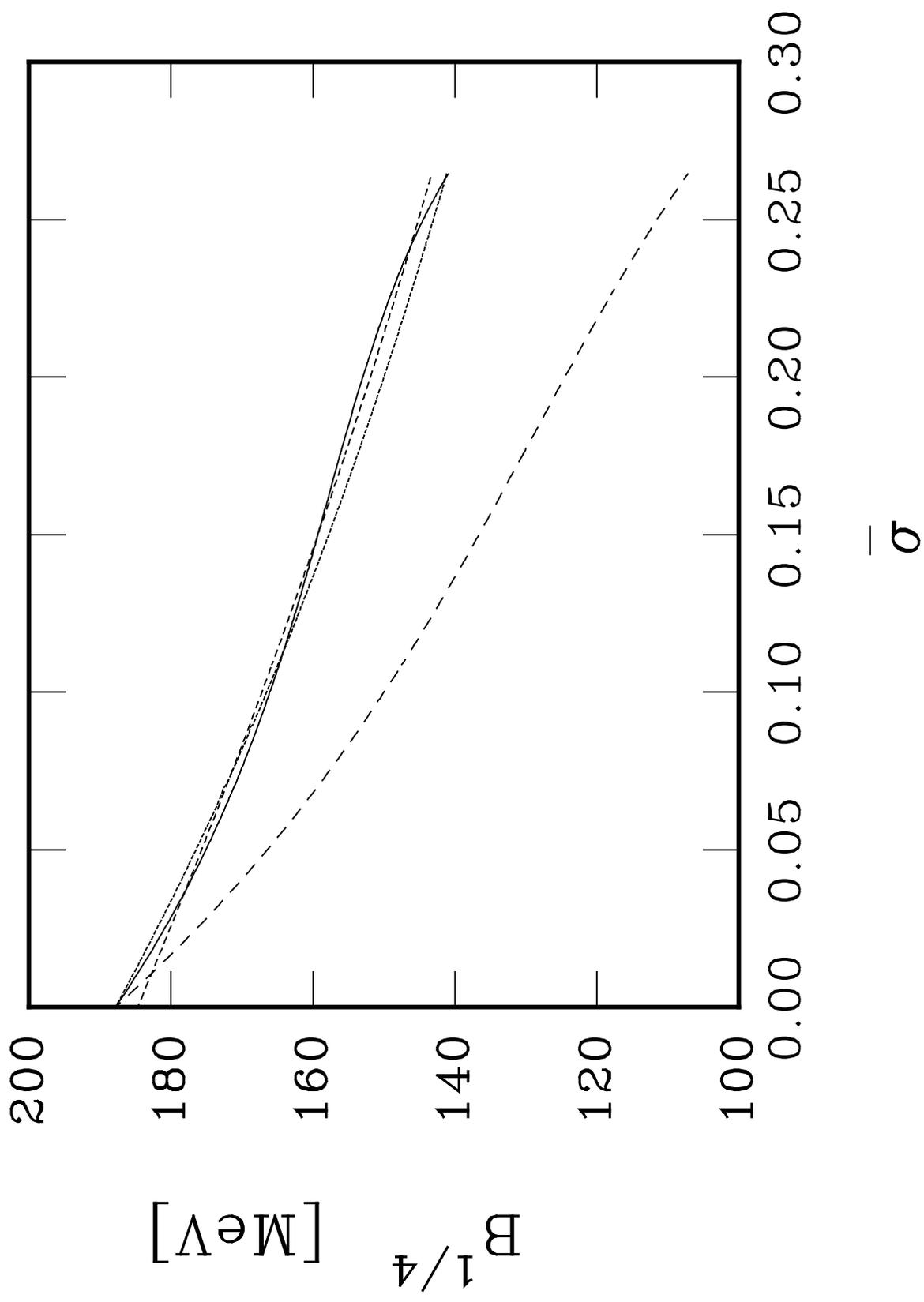
\end{document}